\documentclass{iopart}
\usepackage{graphicx}
\usepackage{epsfig}
\usepackage{iopams}

\begin{document}

\title{On reduced density matrices for disjoint subsystems}
\author{Ferenc Igl\'oi}
\address{Research Institute for Solid
State Physics and Optics, H-1525 Budapest, P.O.Box 49, Hungary}
\address{
Institute of Theoretical Physics,
Szeged University, H-6720 Szeged, Hungary} 
\author{Ingo Peschel}
\address{ Fachbereich Physik, Freie Universit\"at Berlin, Arnimallee
 14, D-14195 Berlin, Germany}

\begin{abstract}
We show that spin and fermion representations for solvable quantum chains
lead in general to different reduced density matrices if the subsystem is 
not singly connected. We study the effect for two sites in XX and XY chains 
as well as for sublattices in XX and transverse Ising chains.

\end{abstract}

\newcommand{\eq}[1]{\begin{equation}#1\end{equation}}
\newcommand{\dd}{\mathrm{d}}
\newcommand{\ee}{\mathrm{e}}

\section{Introduction}

The investigation of entanglement features in many-body quantum systems
has been a topic of intense research in recent years \cite{Amico08}. In such
studies, one divides the system in two parts in space and asks how these
are coupled in the quantum state. This can be answered from the reduced
density matrix (RDM) for one of the subsystems. For a number of solvable lattice
models, these density matrices can be found exactly \cite{Peschel/Eisler09}
and their spectra determine the entanglement entropy which is a simple
measure of the coupling. In the ground state of typical systems with 
short-range interactions, it is connected with the interface between the
two parts and proportional to its extent, although there are corrections
to this ``area law'' for fermions  \cite{Eisert08,CC09}.

Commonly, one divides the system into two parts which are singly connected,
and most results were obtained for this geometry. However, there are a number of
studies which treat multiply connected subsystems. These were investigated
in one dimension for free fermions and bosons within a continuum approach 
\cite{Casini/Huerta09,Markovitch09} and for conformally invariant systems 
\cite{CC09,CCT09}. On a lattice, spin chains have been treated where the 
subsystem consisted of many single spins separated by a certain number of sites 
\cite{Keating06}, of two blocks of spins, again separated  
\cite{Facchi08,Wichterich09}, and of a whole sublattice \cite{Chen06}.
The latter choice was motivated by the search for phase-transition 
indicators while for two blocks one can use the RDM to study the 
entanglement $\it{between}$ them.
In \cite{Keating06,Facchi08} the calculation was carried out by
transforming the transverse Ising (TI) model into fermions and determining
the density-matrix eigenvalues from the fermionic correlation functions  
\cite{Peschel03,Vidal03,Latorre04,Cheong/Henley04}. The basic finding
for isolated spins was that, due to the many ``interfaces'' with the 
surrounding, the entanglement entropy becomes extensive. 

In this note, we want to point out that the reduced density matrices for the
spin representation and the fermionic representation are, in general, not 
identical and that correspondingly also the entanglement entropies differ, 
if one deals with disjoint subsystems. This might be surprising at first, but
is connected with the non-local structure of the Jordan-Wigner transformation. 
Thus, to obtain a transverse spin correlation function, one needs information 
about a whole string of sites in the fermionic picture \cite{LSM61}.
This is not necessary 
if one asks only for fermionic correlations. Thus the two RDM's contain
different information. In the following section 2, we demonstrate this for 
the simple and analytically solvable case of a subsystem of two sites in 
an XX chain. In section 3 we indicate the generalization to the XY chain. The 
possibly more relevant case of sublattice RDM's is treated in section 4 
for XX and TI rings and the results are summarized in section 5.

\section{Two sites in an XX chain}

In the following we consider the spin one-half XX chain described by the Hamiltonian 
\begin{equation}
H= - \frac {1}{2} \sum_m \left[  \sigma^x_m \sigma^x_{m+1}
+ \sigma^y_m \sigma^y_{m+1} \right].  
\label{XX}
\end{equation}
Here the $\sigma^{\alpha}_m$ are Pauli matrices at site $m$. In terms of fermionic
creation and annihilation operators
$c_m^{\dagger},c_m$, $H$ reads \cite{LSM61} 
\begin{equation} 
H=- \sum_m  (c_m^{\dagger} c_{m+1} + c_{m+1}^{\dagger} c_{m}) 
\label{hopping}
\end{equation}
and corresponds to a simple hopping model. If the system forms a ring, one has 
to take care of a boundary term. In the following we always look
at the ground state and choose as subsystem the sites $1$ and $n$.

\subsection{Spin RDM}

The RDM for two spins in XX, XY and TI chains has been discussed in many papers, see
\cite{Osterloh02,Osborne/Nielsen02,Oliveira06a,Oliveira06b,Campos07,Chen07,Stauber/Guinea09}.
It is a $4 \times 4$ matrix and its elements can be expressed as expectation values 
of proper operators at the two sites, i.e. as two-point correlation functions. Since 
the XX ground state has fixed $S^z_{tot}=0$, all entries corresponding 
to a change of this value are zero. As a result, the RDM in the basis 
$|++\rangle,|+-\rangle,|-+\rangle,|--\rangle$ has the simple form
\begin{equation}
 \rho_{\,\mathrm{S}} =  
    \left(  \begin{array} {llll}
      a & & & \\  
      & b & c &  \\
      & c & b &  \\
      & & & a
  \end{array}  \right)   
  \label{rho-XX}
\end{equation}
where $2a+2b=1$ due to $\mathrm{tr}(\rho_{\,\mathrm{S}})=1$. The non-zero matrix elements 
are given by
\begin{eqnarray}
  a & = & \rho_{\,\mathrm{S}}(++,++) = 
        \frac {1}{4} \langle (1+\sigma^z_1) (1+\sigma^z_n) \rangle ,  \\
  b & = & \rho_{\,\mathrm{S}}(+-,+-)= 
        \frac {1}{4} \langle (1+\sigma^z_1) (1-\sigma^z_n) \rangle ,  \\
  c & = & \rho_{\,\mathrm{S}}(+-,-+)= 
        \frac {1}{4} \langle \sigma^{+}_1 \sigma^{-}_n \rangle 
 \label{coeff1}
\end{eqnarray}
where $\sigma^{\pm}=\sigma^x \pm \sigma^y$. In the ground state,  
$\langle \sigma^z_m \rangle = 0$ and the expressions become
\begin{eqnarray}
  a & = & \frac {1}{4} + \frac {1}{4}\langle \sigma^z_1 \sigma^z_n \rangle ,  \\
  b & = &  \frac {1}{4} - \frac {1}{4}\langle \sigma^z_1 \sigma^z_n \rangle ,  \\
  c & = &  \frac {1}{2} \langle \sigma^{x}_1 \sigma^{x}_n \rangle
 \label{coeff2}
\end{eqnarray}
In this way, the RDM is expressed completely in terms of standard spin correlation
functions. This approach has also been used to determine RDM's in the XXZ model, 
see e.g. \cite{Sato/Shiroishi07} and references therein.

The correlation functions can be calculated in the fermionic representation. Then
one has
\begin{equation}
 \frac {1}{4} \langle \sigma^z_1 \sigma^z_n \rangle  =  -\, C_{1,n}C_{n,1} =
 -\, C_{1,n}^2
\label{corr-zz}
\end{equation}
where $C_{i,j}=\langle c^{\dagger}_i c_j \rangle$ is the fermionic single-particle
function. For the $xx$ correlations one finds the determinant \cite{LSM61}
\begin{equation}
 \frac {1}{2} \langle \sigma^{x}_1 \sigma^{x}_n \rangle=2^{n-2}
\left| \begin{array}{llcl}
        C_{1,2} & C_{1,3} & \cdots & C_{1,n}  \\
        \bar C_{2,2} & C_{2,3} & \cdots & C_{2,n}  \\
        C_{3,2} & \bar C_{3,3} & \cdots & C_{2,n}\\
        \dotfill &&&\\\
        C_{n-1,2} & C_{n-1,3} & \cdots & C_{n-1,n}  
        \end{array} \right| 
 \label{corr-xx}
\end{equation}
Here the quantities $\bar C_{m,m}$ are given by $\bar C_{m,m}=C_{m,m}-1/2$  and vanish 
in the ground state, for which the fermionic system is half filled. In two cases
the expression reduces to a single term, namely if the two sites are nearest 
neighbours ($n=2$), and if they are the ends of an open chain. Then the effect of the
Jordan-Wigner strings disappears and
\begin{equation}
   \frac {1}{2} \langle \sigma^{x}_1 \sigma^{x}_n \rangle  =  C_{1,n}
 \label{corr-xx1}
\end{equation}
The ``long-distance entanglement'' in the latter case has been studied e.g. in 
\cite{Campos07,Giampaolo09}.

\subsection{Fermion RDM}

In the fermionic case, one can proceed in exactly the same way. The RDM is
again a $4 \times 4$ matrix, this time in the basis specified by the occupation
numbers $|11\rangle,|10\rangle,|01\rangle,|00\rangle$. Since the ground state
has fixed total particle number $\cal{N}$, all elements corresponding to a 
change of $\cal{N}$ vanish. Therefore $\rho_{\,\mathrm{F}}$ has again the 
form (\ref{rho-XX}) and the elements are now given by
\begin{eqnarray}
  a & = & \langle  c_1^{\dagger} c_1 c_n^{\dagger} c_n \rangle = 
        \frac {1}{4} - C_{1,n}^2  ,  \\
  b & = & \langle  c_1^{\dagger} c_1(1- c_n^{\dagger} c_n) \rangle =                
        \frac {1}{4} + C_{1,n}^2  ,  \\
  c & = & \langle  c_1^{\dagger} c_n \rangle =  C_{1,n}.  
  \label{coeff3}
\end{eqnarray}
One sees that $a$ and $b$ are the same as in the spin representation, but
$c$ is in general different. Instead of the $xx$ spin correlation function, 
the fermionic one-particle correlation function $C_{1,n}$ appears. This is the basic 
difference between the two representations. Only for the two exceptional cases
mentioned above, the expressions for $c$ coincide. Then the
two spins are nearest neighbours, or can be considered as neighbours by bending
the open chain to a ring, and correspondingly the subsystem is singly connected.

The appearance of $C_{1,n}$ is natural, since with $\rho_{\,\mathrm{F}}$ one
must in turn be able to calculate this correlator. That essentially only this
quantity enters, could have been seen also from the general result
for fermionic RDM's \cite{Peschel03,Peschel/Eisler09} 
\begin{equation}
 \rho_{\,\mathrm{F}} =  \frac{1}{Z}\; \exp{(-\cal{H})}
  = \frac{1}{Z}\; \exp{(-\sum_{i,j=1,n} H_{i,j} c_i^{\dagger} c_j )}
\label{rho-F}
\end{equation}
where $Z$ ensures $\mathrm{tr}(\rho_{\mathrm{\,F}})=1$ and 
the matrix $H_{i,j}$ in the exponent follows from the correlation matrix 
involving the sites of the subsystem, in our case $C_{1,1}=C_{n,n}=1/2$
and $C_{1,n}$. One can obtain the form (\ref{rho-XX}) also from (\ref{rho-F}), 
but the route taken above is much simpler.\\

\subsection{Entanglement entropies}

>From the structure of $\rho_{\,\mathrm{S}}$ and $\rho_{\,\mathrm{F}}$ one reads off 
the four eigenvalues
\begin{equation}
   w_1=b+c, \,\,\,\, w_2 = w_3 = a, \,\,\,\, w_4 =b-c. 
  \label{eigenvalues}
\end{equation}
The entanglement entropy is then given by $S=-\sum  w_k\ln w_k$.
Since the correlations go to zero for large separations, all $w_k$ approach $1/4$ 
in this limit and $S$ goes to the value $S= 2\ln2$ in both representations.
This is the sum of two single-spin contributions and also the maximum one can have. 

\begin{figure}[t]
\begin{center}
\includegraphics[width=9.cm,angle=0]{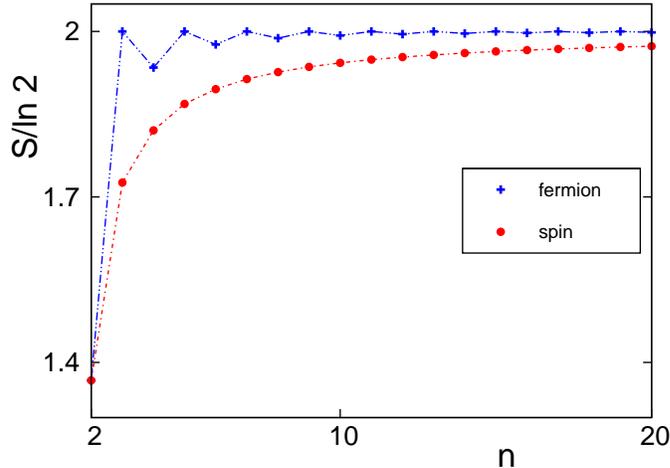}
\end{center}
\caption{Two-site entanglement in the infinite XX chain as a function of the distance $n$, 
calculated in the spin and in the fermion basis.
\label{fig1}
}   
\end{figure}

The complete behaviour of $S$ is obtained by using the correlation function of the 
infinite chain
\begin{equation}
 C_{i,j}= \frac {\sin(\pi(i-j)/2)}{\pi(i-j)}
\label{corrXX}
\end{equation}
and calculating the eigenvalues $w_k$ numerically. The result is shown in Fig.\ref{fig1}. 
According to the previous remarks, the two entropies coincide for $n=2$ and for
$n \rightarrow \infty$. In between, the spin entropy 
$S_{\,\mathrm{S}}$ always lies below the fermionic one, $S_{\,\mathrm{F}}$, because the
$xx$-correlations only decay as $n^{-1/2}$. If the distance $n-1$ is even, i.e. if the
two sites are on the same sublattice, the correlation $ C_{1,n}$ vanishes and 
$S_{\,\mathrm{F}}$ has the maximum value $2\ln2$. Apart from this feature, the  
asymptotic behaviour is determined by $c$
\begin{equation}
   S \simeq 2\ln2 - 4 c^2
  \label{asymp}
\end{equation}
which gives a $1/n$ approach to the limit in the spin picture and a $1/n^2$ 
variation in the fermion picture.\\


\section{Anisotropic chains}

The previous considerations can be generalized to the case of an XY or a
TI chain. Then the rotational symmetry is absent, but the ground state is
built from configurations with either an even or an odd number of (+) spins.
Then 
\begin{equation}
 \rho_{\,\mathrm{S}} =  
    \left(  \begin{array} {llll}
      a & & & d \\  
      & b & c &  \\
      & c & b &  \\
      d & & & a
  \end{array}  \right)   
  \label{rho-XY}
\end{equation}
and the additional matrix element $d$ is given by
\begin{equation}
  d  =  \rho_{\,\mathrm{S}}(++,--)= 
        \frac {1}{4} \langle \sigma^{+}_1 \sigma^{+}_n \rangle
 \label{coeff4}
\end{equation}
In the fermionic picture, on the other hand
\begin{equation}
  d  =  \rho_{\,\mathrm{F}}(11,00)= 
        \langle c^{\dagger}_1 c^{\dagger}_n \rangle
 \label{coeff5}
\end{equation}
Again, both expressions are different, except for $n=2$ or for end spins. 
The presence of $d$ splits the two degenerate eigenvalues of $\rho$ and gives 
$w_{2,3}=a \pm d$.

An additional feature arises from the long-range order which exists in the anisotropic 
XY model and in the TI model with strong coupling. In the spin picture, the coefficients 
$c$ and $d$ do not vanish for large separations and therefore the entropy does not become 
the sum of two single-site contributions. If $\gamma$ denotes the anisotropy of the XY 
model with zero field, the asymptotic values are \cite{McCoy68}
\begin{equation}
  c  =  d = \frac {1}{2}  \langle \sigma^{x}_1 \sigma^{x}_n \rangle =
            \frac {1}{2} \frac {\sqrt\gamma}{1+\gamma}
 \label{xy}
\end{equation}
and vary between $0$ for $\gamma=0$ and $1/4$ for $\gamma=1$. Correspondingly,  
$S_{\,\mathrm{S}}$ varies between $2\ln2$ and $\ln2$.
The fermionic correlation functions, on the other hand, approach zero for large
separations, as in the XX model, and  $S_{\,\mathrm{F}}= 2\ln2 $ independent of $\gamma$. 
Thus the quantities $S_{\,\mathrm{S}}$ and $S_{\,\mathrm{F}}$ have different
asymptotic values. Only for end spins in open chains, the asympotic values coincide
and are in this case related to the surface order. This happens also in dimerized XX 
chains \cite{Campos07,Giampaolo09}.

\section{Sublattice entanglement}

We now turn to the case where the subsystem consists of every second site in a chain.
For the infinite XX model, it then follows from (\ref{corrXX}) that the fermionic
correlations vanish for unequal sites on the same sublattice. This holds also for finite
rings, as well as for non-homogeneous couplings. Therefore the correlation matrix $C_{i,j}$ 
on one sublattice is diagonal, $C_{i,j}=1/2 \;\delta_{i,j}$. The RDM obtained via  
(\ref{rho-F}) is then also diagonal, all eigenvalues $w_k$ are equal and 
$S_{\,\mathrm{F}} = L \ln2$ if the system is a ring with $N=2L$ sites. This is the extensivity 
mentioned in the introduction, and the result has a simple interpretation. The ground state 
$|\Psi\rangle$ is obtained by filling the single-particle eigenstates of (\ref{hopping}) for 
momenta $|q| <  \pi/2$. But the corresponding operators can be decomposed as
\begin{eqnarray}
c^{\dagger}_q= \frac {1}{\sqrt{N}} \sum_{m=1}^L \exp{(-iq2m)}c^{\dagger}_{2m}
            + \frac {1}{\sqrt{N}} \sum_{m=1}^L \exp{(-iq(2m+1))}c^{\dagger}_{2m+1} \nonumber \\
            = \frac {1}{\sqrt2}(a^{\dagger}_q+b^{\dagger}_q) 
\label{sublattice1}
\end{eqnarray}
Therefore  
\begin{equation}
 |\Psi\rangle  =  \prod_{|q|<\pi/2}\frac {1}{\sqrt2}(a^{\dagger}_q+b^{\dagger}_q)|0\rangle  
\label{sublattice2}
\end{equation}
is a product of fermionic triplets formed from modes with the same $q$ on different 
sublattices. Each of them contributes $\ln2$ to the entanglement entropy. \\

\begin{figure}[htb]
\begin{center}
\includegraphics[width=6.3cm,angle=0]{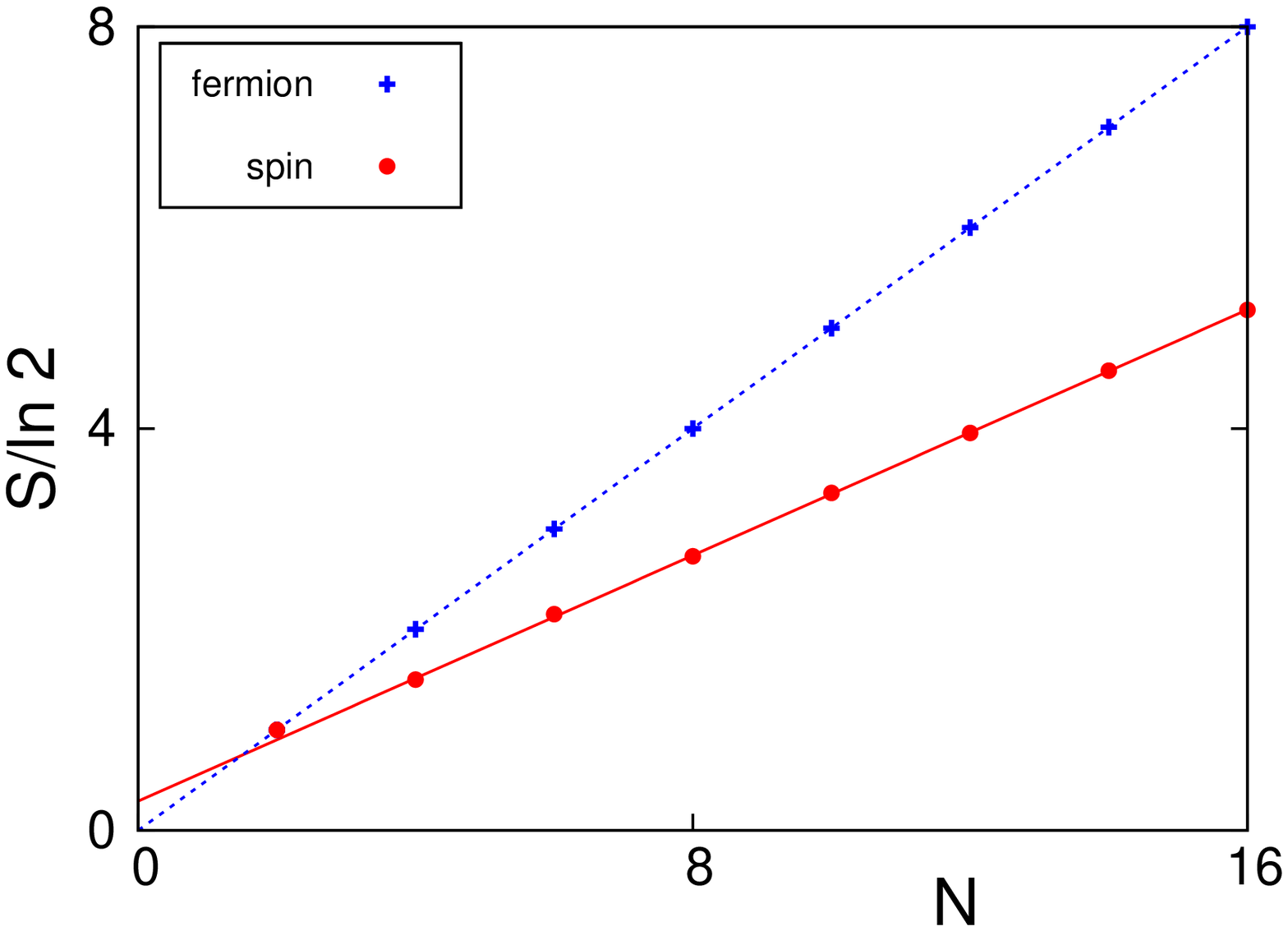}
\includegraphics[width=6.3cm,angle=0]{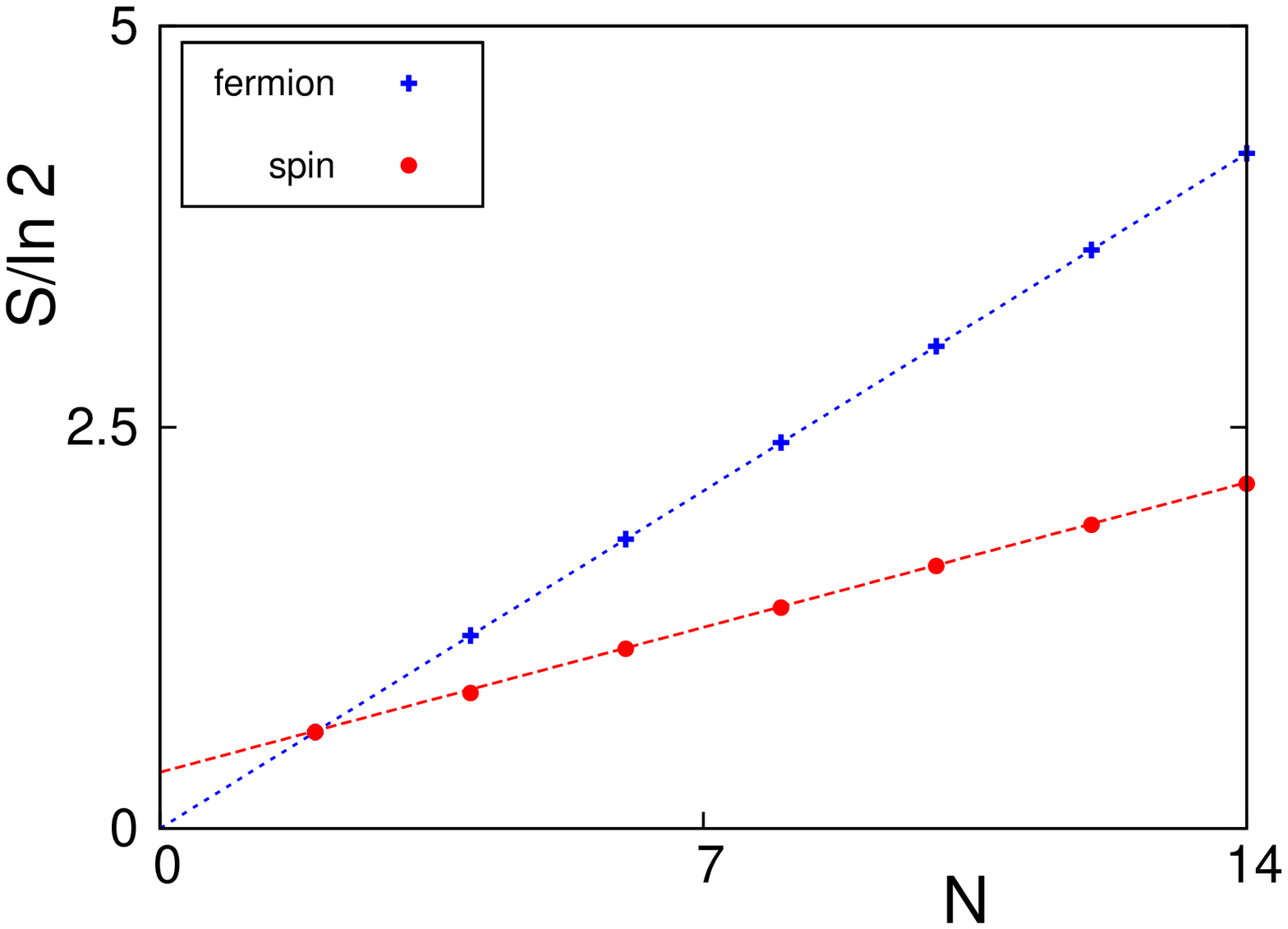}
\caption{Sublattice entanglement entropy for rings of length $N=2L$, calculated
in the spin and in the fermion basis. Left: XX model, right: critical TI model.
The straight lines represent the respective asymptotic behaviour, see the text.}
\label{fig2} 
\end{center}  
\end{figure}

For the spin RDM, a calculation for $N=4$ already shows that the $w_k$ are not
all equal, but given by $w_1=1/2, w_2=w_3=1/4, w_4=0$. The corresponding entropy
is $S_{\,\mathrm{S}}=3/2 \ln2$ and thus smaller than $S_{\,\mathrm{F}}=2 \ln2$.
To investigate the size dependence, we have calculated  $\rho_{\,\mathrm{S}}$ numerically
by finding the ground state of $H$ in the subspace $S^z_{tot}=0$ for $N$ up to 16 sites.
The resulting values are plotted in Fig.\ref{fig2} on the left.\\


The data can be fitted for the larger $L$ by $S_{\,\mathrm{S}}(L)/\ln2=s*L+s_0$ with 
$s=0.6116(4)$ and $s_0=0.290(5)$. In addition, there are subleading corrections which are 
different for even and odd values of $L/2$, related to the slightly alternating behaviour of 
$\rho_{\,\mathrm{S}}$. Thus the extensivity of $S$ holds also in the spin picture, but the 
value is reduced to about 60\% of the fermionic one.
For open chains, the values for $S_{\,\mathrm{S}}$ lie about 0.5 higher than for the rings,
but the slope $s$ is similar.  \\

We have also considered the XX chain with enforced dimerization where the coupling between
sites $m$ and $m+1$ is given by $J_m=1+\delta \, (-1)^m$. Then one has a finite correlation
length if $\delta \neq 0$. However, the fermionic sublattice entanglement is not affected. 
The expectation values $\langle c^{\dagger}_i c_j \rangle$ still vanish on the same 
sublattice and one can interpret the result $S_{\,\mathrm{F}} = L \ln2$ as before. The spin
entanglement, on the other hand, is an even function of $\delta$ and depends on the 
dimerization. In particular for $\delta = \pm 1$, where the system decomposes into coupled 
pairs of sites each of which contributes $\ln2$, it becomes equal to $S_{\,\mathrm{F}}$.


\begin{figure}[h]
\begin{center}
\includegraphics[width=8.5cm,angle=0]{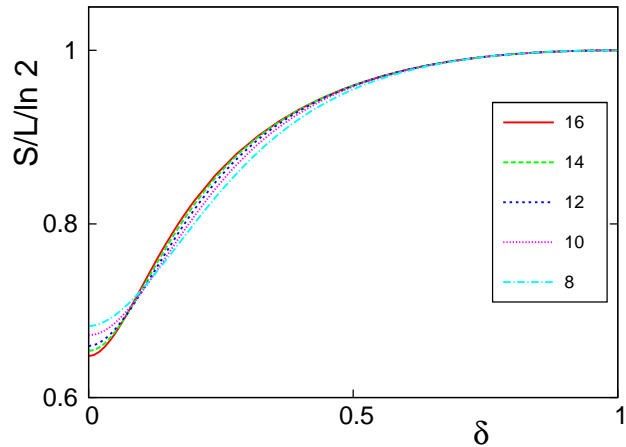}
\end{center}
\caption{Sublattice entanglement entropy per site of XX rings as a function
of the dimerization, calculated in the spin basis for different sizes $N$.}
\label{fig3}  
\end{figure}


The detailed variation with $\delta$ for fixed $N$ is shown in Fig.\ref{fig3}. Since
$S/L$ is plotted, the large-$L$ result is basically the slope $s$ introduced above and
seen to increase monotonously from 0.6116 to 1 as  $\delta$ varies between 0 and 1.
Near $\delta=0$ one finds a special feature due to an eigenvalue 
$w_k \sim \delta^2$ of  $\rho_{\,\mathrm{S}}$ which leads to a non-analytic dependence 
$S_{\,\mathrm{S}}(L) \sim \delta^2\ln |\delta|$.\\

Finally, we have investigated the transverse Ising model on a ring with 
Hamiltonian
\begin{equation}
H= -  \sum_{m=1}^{N} \left[  \sigma^x_m \sigma^x_{m+1}
+ h \,\sigma^z_m  \right].  
\label{TI}
\end{equation}
and calculated the sublattice entanglement in both representations. The result for the
critical case $h=1$ is very similar to that for the XX model and shown in 
Fig.\ref{fig2} on the right for up to $N=14$ sites. The curves can again be fitted by a 
linear function and give $s=0.258(2)$ and $s_0=0.700(1)$ in the spin picture while
$S_{\,\mathrm{F}}$ is exactly proportional to $L$ with a slope $s=0.6001$. Thus both
entropies are extensive but the fermionic value is more than twice the spin value.

More interesting is the non-critical case. Fig. 4 shows on the left the results for
the spin representation. One sees that for the larger sizes,  $S_{\,\mathrm{S}}/L$ shows a
a maximum near the criticial value $h=1$. In the disordered region $h>1$, the curves
approach a limit and $S$ is therefore extensive, but in the ordered region $h<1$, they
depend on $L$ and $S$ is not extensive. Rather one has $S_{\,\mathrm{S}}\rightarrow \ln2$ 
for $h \rightarrow 0$. This is the same value as for a subsystem in the form of one block
and has the same origin. The ground state in this limit is a superposition of the two
states with all spins having $\sigma^x=+1$ and all having $\sigma^x=-1$, respectively. 
This GHZ state leads to a RDM with two non-zero eigenvalues $w=1/2$ and thus to $\ln2$ 
for the entropy.

%
%
\begin{figure}[h]
\begin{center}
\includegraphics[width=6.3cm,angle=0]{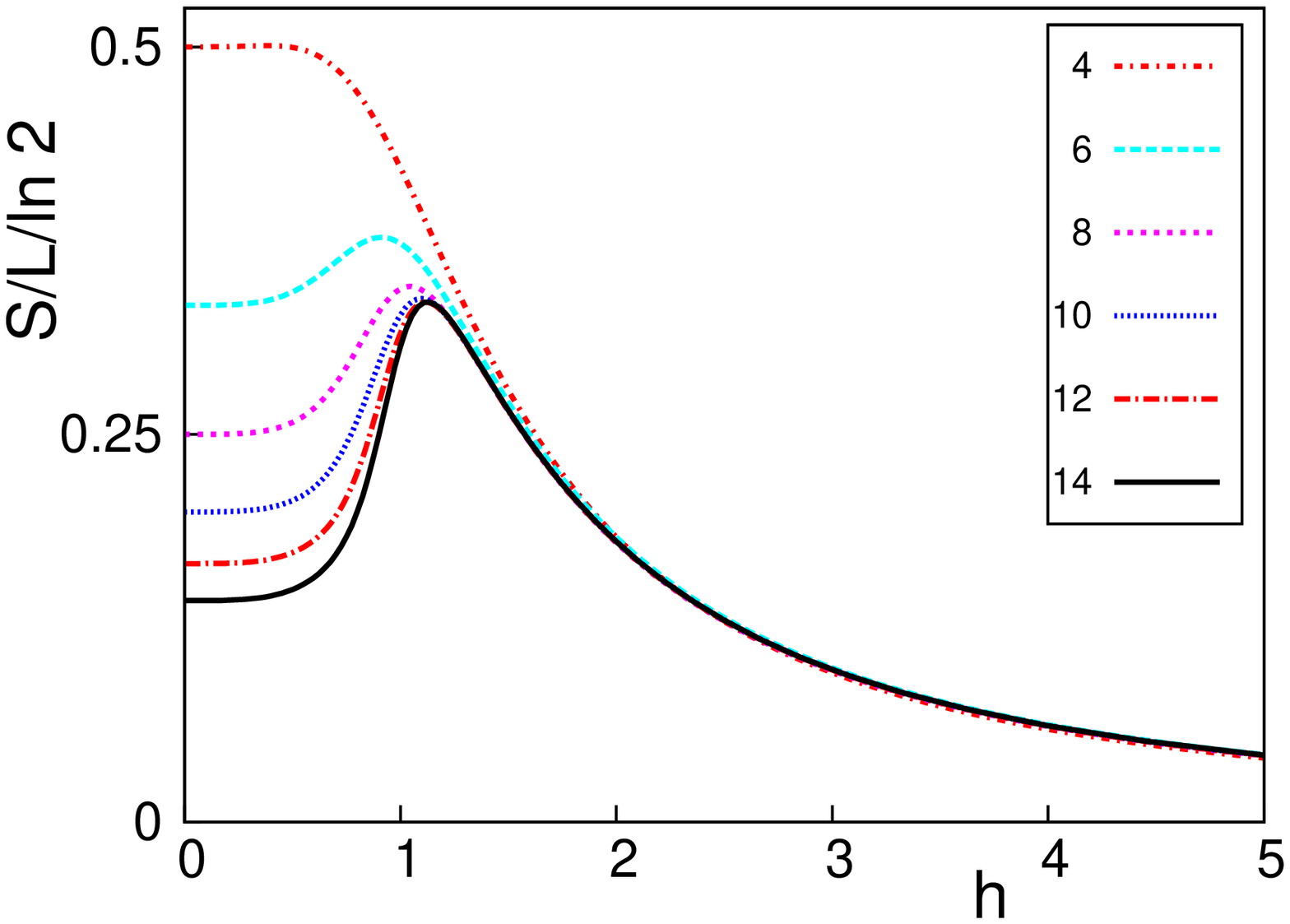}
\includegraphics[width=6.3cm,angle=0]{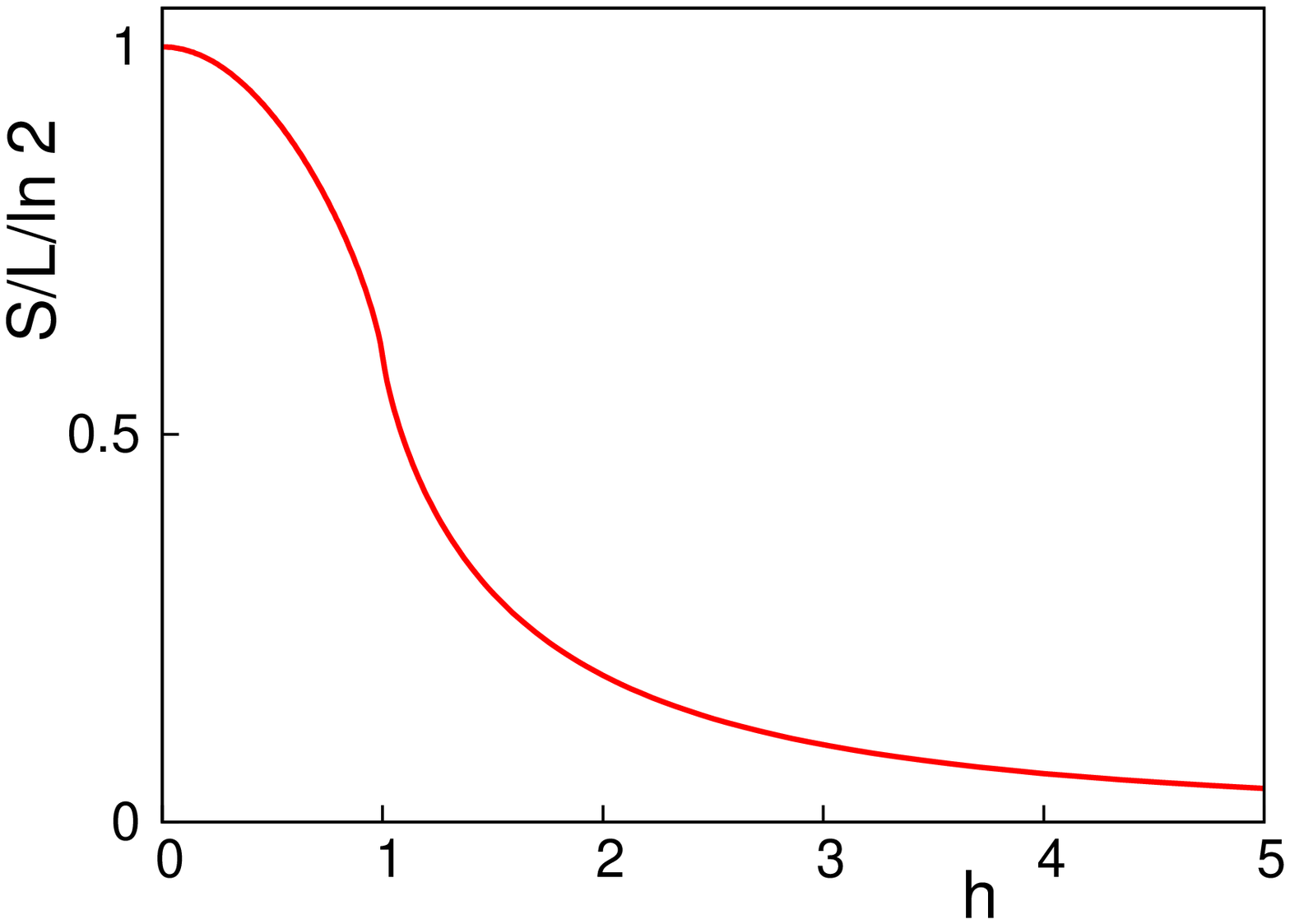}
\end{center}
\caption{Sublattice entanglement entropy per site in TI rings
as a function of the transverse field. 
Left: calculated in the spin basis for rings of different size $N$. 
Right: calculated in the fermion basis in the large-$N$ limit.
Note the different vertical scales.} 
\label{fig4}   
\end{figure}
%
%
The fermionic case is different and can be treated analytically, although the ground 
state is more complicated than in the XX model. Following 
\cite{Peschel03,Peschel/Eisler09}, one has to find the eigenvalues $(2\zeta-1)^2$ of 
the matrix  
\begin{equation}
 {\bf{M}}= (2{\bf{C}}-{\bf{1}}-2{\bf{F}})(2{\bf{C}}-{\bf{1}}+2{\bf{F}})
\label{pair}
\end{equation}
where  $F_{i,j}=\langle c_i^{\dagger} c_j^{\dagger}\rangle$ and all sites are on one
sublattice. However, due to the translational invariance of the subsystems, one can 
work in momentum space. Then the  matrix ${\bf{M}}$ decomposes into $2 \times 2$ blocks 
involving the correlation functions $C_q=\langle a_q^{\dagger} a_q \rangle$, 
$C_{-q}=\langle a_{-q}^{\dagger} a_{-q} \rangle$ 
and $F_q=\langle a_q^{\dagger} a_{-q}^{\dagger} \rangle$. As a result, one finds
\begin{equation}
 (2 \zeta_q-1)^2 = \frac {1}{2} \left[ 1+\frac {h^2-1}
                    {[(1+h^2)^2-4h^2 \cos^2q]^{1/2}} \right]
\label{zeta-q}
\end{equation}
from which the entanglement entropy is obtained as 
\begin{equation}
 S_{\,\mathrm{F}}=-\sum_{0<q<\pi/2} \left[ \zeta_q\ln\zeta_q+(1-\zeta_q)\ln(1-\zeta_q) \right]
\label{zeta-ent}
\end{equation}
This is always extensive and shown in Fig. 4 on the right. For $h \gtrsim 2.5$ the curve
practically coincides with the one in the spin representation. This can be attributed
to the short correlation length whereby only neighbouring sites see each other.
The asymptotic form is $S/L \sim \ln h/h^2$ since $\zeta_q=1-O(1/h^2)$ for large $h$.     
As $h$ is reduced, however, the curve rises continuously to one, corresponding to the 
value $S_{\,\mathrm{F}}=L\ln2$, with no sign of the long-range order.
For three values of $h$, the $\zeta_q$ are independent of $q$, namely
for $h=0$ ($\zeta_q=1/2,S_{\,\mathrm{F}}=L\ln2 $), for $h=\infty$ 
($\zeta_q=1,S_{\,\mathrm{F}}=0$) and for $h=1$ where $\zeta_q=(1+1/\sqrt2)/2$ and 
\begin{equation}
  S_{\,\mathrm{F}}= L \;\left[ \frac {3}{2} \ln2 - \frac {1}{\sqrt2} \ln(1+\sqrt2) \right]
\label{critentropyTI}
\end{equation}
This gives the value $s=0.6001$ cited above. Near $h=1$, the entropy per site varies
asymptotically as $(h-1)\,\ln|h-1|$ and thus has infinite slope. This is the
signature of the phase transition in $S_{\,\mathrm{F}}$.
The interpretation of the extensivity is similar as in the XX model, but in the ground 
state
\begin{equation}
 |\Psi\rangle  =  \prod_{0<q< \pi}\left[u_q +v_q c^{\dagger}_qc^{\dagger}_{-q} \right]|0\rangle 
\label{groundTI}
\end{equation}
one has now a coupling of the $\it{two}$ single-particle states $(q,-q)$ in the $\it{two}$
sublattices if one inserts (\ref{sublattice1}). The corresponding operators appear twice
in the product, since $c^{\dagger}_{\pi-q} =(a^{\dagger}_{-q}-b^{\dagger}_{-q})/ \sqrt2$.

Since TI and XX chains are related by a dual transformation, one could expect a relation 
between the corresponding entanglements. If the subsystem is a block in a ring, such a 
connection indeed exists \cite{Igloi/Juhasz08}. In the sublattice case, however, we have 
not found a similar result.

\section{Conclusion}

We have studied the entanglement in spin chains for the case that the subsystem
is not singly connected. We demonstrated that working in the spin and the fermion
representation leads in general to different RDM's and to different entanglement
entropies. We did this by looking at two extreme cases, namely a subsystem of only
two sites and one in the form of a whole sublattice. The first one displays the
effect particularly clearly, while the second one provides an example, where it
is particularly large.

At the level of the wave function, there is no difference between the two 
representations. One can rewrite the spin expression directly into the occupation-number
form. However, the operators sample different information. In the fermion picture,
the spin function  $\langle \sigma^{x}_i \sigma^{x}_j\rangle$ needs all sites between 
$i$ and $j$, whereas in the spin picture the same is true for the fermion function  
$\langle c^{\dagger}_i c_j \rangle$. If some of these sites do not belong to the
chosen subsystem, the reduced density matrices giving these expectation values will
usually not coincide. If one determines them by integrating out degrees of freedom, 
this difference arises, because one needs Grassmann variables in the fermionic case
\cite{Chung/Peschel01,Cheong/Henley04}. This can lead to sign changes in the terms
contributing to a particular matrix element, as compared to the spin calculation, and thus 
to a different final result. One can see this explicitly by considering an XX chain with 
four sites. Thus while the eigenvalues of the spin RDM are directly related to the
coefficients in the Schmidt decomposition of the state, this does not necessarily 
hold for those of the fermion RDM. They and the resulting entropy measure the 
entanglement in a somewhat different way.\\ 

$\it{Note\;\, added.}$ The difference between the two representations discussed here
has also been noted in a preprint by V. Alba et al., arXiv:0910.0706, which just 
appeared.\\

\ack{We thank H. Wichterich for drawing our attention to the problem and for 
correspondence.This work has been supported by the Hungarian National Research Fund 
under grant No OTKA K62588, K75324 and K77629 and by a German-Hungarian exchange program 
(DFG-MTA). }

\end{document}